\documentclass[sigconf,authordraft,screen,nonacm,review=false,timestamp=false]{acmart}

\usepackage{makecell}
\usepackage{float}
\usepackage{booktabs}
\usepackage{arydshln}
\usepackage{ulem}
\usepackage{threeparttable}
\usepackage{makecell}

\definecolor{dkgreen}{rgb}{0,0.6,0}
\definecolor{gray}{rgb}{0.5,0.5,0.5}
\definecolor{mauve}{rgb}{0.58,0,0.82}

\newcommand{\textbfm}[1]{\textbf{\textcolor{blue}{#1}}}
\newcommand{\textbfg}[1]{\textcolor{red}{#1}}

\AtBeginDocument{%
  \providecommand\BibTeX{{%
    \normalfont B\kern-0.5em{\scshape i\kern-0.25em b}\kern-0.8em\TeX}}}

\setcopyright{acmlicensed}
\copyrightyear{2024}
\acmYear{2024}
\acmDOI{XXXXXXX.XXXXXXX}

\acmConference[KDD'24]{the 30th ACM SIGKDD Conference on Knowledge Discovery and Data Mining}{August 25--29,
  2024}{Barcelona, Spain}
\acmISBN{978-1-4503-XXXX-X/18/06}

\begin{document}

\title{Social Links vs. Language Barriers:\\ Decoding the Global Spread of Streaming Content}

\author{Seoyoung Park}
\authornote{Both authors contributed equally to this research.}
\author{Sanghyeok Park}
\authornotemark[1]
\affiliation{%
  \institution{Soongsil University}
  \city{Seoul}
  \country{Republic of Korea}
}
\email{sypark19@soongsil.ac.kr}
\email{ps2575@soongsil.ac.kr}

\author{Taekho You}
\authornote{Corresponding Author} 
\affiliation{%
  \institution{Pohang University of Science and Technology}
  \city{Pohang}
  \country{Republic of Korea}
}
\email{th_you@postech.ac.kr}

\author{Jinhyuk Yun}
\authornote{Corresponding Author}
\affiliation{%
  \institution{Soongsil University}
  \city{Seoul}
  \country{Republic of Korea}
}
\email{jinhyuk.yun@ssu.ac.kr}

\renewcommand{\shortauthors}{Park and Park, et al.}

\begin{abstract}
The development of the internet has allowed for the global distribution of content, redefining media communication and property structures through various streaming platforms. Previous studies successfully clarified the factors contributing to trends in each streaming service, yet the similarities and differences between platforms are commonly unexplored; moreover, the influence of social connections and cultural similarity is usually overlooked. We hereby examine the social aspects of three significant streaming services--Netflix, Spotify, and YouTube--with an emphasis on the dissemination of content across countries. Using two-year-long trending chart datasets, we find that streaming content can be divided into two types: video-oriented (Netflix) and audio-oriented (Spotify). This characteristic is differentiated by accounting for the significance of social connectedness and linguistic similarity: audio-oriented content travels via social links, but video-oriented content tends to spread throughout linguistically akin countries. Interestingly, user-generated contents, YouTube, exhibits a dual characteristic by integrating both visual and auditory characteristics, indicating the platform is evolving into unique medium rather than simply residing a midpoint between video and audio media.
\end{abstract}





\maketitle

\section{Introduction}
We live in an era of globalization. People now share their culture in real-time; it is no longer limited to the local area. For a long time, one has considered that cultural content mainly spread through the movement of people, while the increasing migration of these media to the online realm is one of the most prominent features of the Internet age of the twenty-first century~\cite{park2019research}. Online streaming platforms enable us to exhibit their content through the internet without physical (or geographical) barriers; thus, the limitation of direct human mobility on the spread of content has significantly diminished. This transition has changed the landscape of content consumption. For example, the promotion of recorded music has changed from the conventional purchase of albums and singles in different physical formats to digital formats via the Internet~\cite{brown2016buy}, changing the concept of psychological ownership in music streaming consumption~\cite{sinclair2017psychological}. Similarly, the traditional way of consuming movies in theaters has shifted to the convenience of watching the latest films anytime, anywhere through over-the-top (OTT) platforms, erasing spatial constraints~\cite{mulla2022assessing}.

However, the spreading of content is not, nevertheless, affected only by physical barriers. Consumption and creation of cultural content are also influenced by historical events and personal preferences~\cite{michel2011quantitative}. The presence of shared cultural traits and divergent cultural characteristics between countries can either help or restrain the dissemination of specific types of content~\cite{baek2015relationship}. Linguistic affinities between two groups facilitate the dissemination of information and cultural exchange~\cite{lazear1999culture}. On the other hand, contemporary information technology provides interactive online platforms, \textit{e.g.} social networks and internet messengers, that facilitate the sharing of knowledge. Hence, the advent of the information society gives rise to thought-provoking inquiries: do social interactions impact the spreading of cultural content? If so, how much more of an impact does social media have than linguistic barriers? Does this effect remain the same regardless of the platform or kind of content, or does it vary? However, the fact that many previous researches have concentrated on particular platforms and content kinds limits its possibility of addressing these issues~\cite{sinclair2017psychological, baek2015relationship, mulla2022assessing, duenas2023structure}.

In this study, we use three well-known streaming services--Netflix, Spotify, and YouTube--to try and provide answers to the above issues. While prior research has made significant progress in comprehending cultural dissemination in the online era, additional data with broader coverage is still required, ranging from hours-long films to seconds-long short videos. This dataset selection allows us to examine the spread of culture among various nations. We then employed two cross-country connections to examine the diffusion patterns on these platforms: linguistic similarity between the two countries and social networks. Using these datasets, we are able to determine that the impact of linguistic similarity and social networks on the dissemination of content differs by platform and data. Linguistic similarity significantly influences the dissemination of long video content, as evidenced by the case of Netflix-represented video media, while Spotify's music is disseminated more frequently between two socially interconnected countries, irrespective of language barriers. Conversely, regarding YouTube, we observe that it exhibits distinct attributes compared to the aforementioned platforms—namely, a propensity to consume content generated by users (as opposed to relying on expert groups for content as in the case of Netflix and Spotify) and user-generated content as in the case of YouTube, where users simultaneously serve as providers and consumers.

\section{Related Works}
As internet-based streaming services gradually replaced conventional media services, researchers are beginning to show interest in these streaming services. Various facets of the streaming services were studied, which our study is grounded: media~\cite{gaustad2019streaming, hesmondhalgh2021music}, data science~\cite{cha2007tube, platt2015international, lotz2021between, ibrus2023quantifying}, business and marketing~\cite{carroni2020business, naveed2017co, burroughs2019house, vonderau2019spotify}; thus, in this section, we provide a brief review of the relevant topics and debates related to the social perspective of streaming platforms that we focused.

Some studies considered the factors influencing user engagement and the popularity of content focused on a single streaming platform~\cite{lewis2013content}. For example, how recommendation algorithms on Netflix shape viewing patterns were investigated~\cite{gurmericc2019behavioral}, while the other study surveyed how user playlists and social curation influence music discovery on Spotify~\cite{park2021social}. Scholars were also interested in the modeling of the population dynamics of your-generated content on YouTube~\cite{hoiles2017engagement}. Another study explored the role of content attributes like genre, release date, and production value in determining a show's success on Korean streaming media~\cite{jang2021movie}. Similarly, Park et al.~\cite{park2019global} analyze how music characteristics like tempo and mood influence listener preferences on Spotify. These studies offer a starting point for comprehending each internal dynamics of streaming platforms and are crucial for comprehending how content types affect different audiences, yet lack the consideration of the difference between various streaming platforms.

The impact of social networks on content consumption has also been frequently studied. A study shows how user decisions are influenced by shared viewing experiences and social recommendations~\cite{bakshy2012social}. Similarly, one explored how social playlists on Spotify contribute to music discovery and taste formation by neural collaborative filtering~\cite{girsang2021neural}. Another study examines the impact of memes and social media conversations on the virality of content~\cite{berger2012makes}. These studies provide important insights into how social connections affect content popularity and enable the dissemination of content among various groups, but they do not take into account the cultural context, including language barriers. In addition to its increasing prevalence in some media industries, streaming is also acknowledged as a developing notion of media convergence~\cite{spilker2020dimensions}. Consequently, there was a need for cross-border transmission study.

Many scholars believe that an influential factor for cross-cultural relationships is cultural similarity and difference~\cite{baek2015relationship}. Socioeconomic variables such as gross domestic product, are possible factors to explain the co-consumption; yet, cultural, regional, and historical factors play a primary role~\cite{duenas2023structure}, while socioeconomic factors downplay the spread of content. Furthermore, although online social media enhances global accessibility to cultural products, technological advancements like internet penetration do not lead to a universal convergence of cultures~\cite{park2017cultural}. A study suggests that cultural similarities between countries can influence preferences for specific genres and themes in movies and TV shows~\cite{la2005multiple}. Repeated reports suggest that linguistic and geographical distance~\cite{way2020local, Bello2021Cultural, jang2023global, terroso2023music} affect cross-cultural relationships. These factors and patterns are observed on diverse platforms or content types~\cite{taneja2016global, liew2022network}. Linguistic barriers may diminish using the translation, and so the effectiveness of subtitling and dubbing strategies in making content accessible to international audiences was also examined~\cite{borell2000subtitling}. In short, linguistic similarity can act as a barrier or facilitator for the international spread of video content, yet their relative importance compared with the social connection, especially considering the differences among the platforms and content types, is rarely investigated.

Our work is at the intersection of these pioneers, highlighting the unique aspect of examining social connections alongside language barriers and content types for understanding the global spread of streaming content.

\section{Methods}
\subsection{Collecting online streaming chart data}
We collected trending charts from three global streaming services: i) YouTube trending videos, ii) Spotify daily top chart, and iii) Netflix weekly top chart. We only gathered the data from 10 countries: Brazil, Canada, France, Germany, India, Japan, Mexico, South Korea, the United Kingdom, and the United States, which are available in all three services for consistency.

First, we used the YouTube trending video dataset, retrieved March 20, 2023, from Kaggle~\cite{youtube_2023}, which includes everyday records of the top 200 trending videos for every country. In our target period, from August 12, 2020, to February 28, 2023, there are 1,820,130 records total in the dataset, which includes 262,721 distinct videos. In addition, topical categories were collected for each video using the YouTube Data API (\url{https://developers.google.com/youtube/v3}; elements of \texttt{topicDetails.topicCategories}) on December 20, 2023; although there also are topical categories in the Kaggle dataset, we self-collected to enhance the accuracy because there are no detailed descriptions of the categories in the Kaggle dataset. Note that there are removed or unlisted videos on YouTube at the API data collection, and thus YouTube API response only 250,186 distinct videos with 1,747,670 records; for the categorical analysis in Section~\ref{subsec:categories}, we only used the videos that can retrieve the category information.

We also collected the Spotify daily top charts between November 7 and November 8, 2023, from the official website (\url{https://charts.spotify.com}). The dataset contains daily records of the top 200 tracks for each country, composed of 23,738 unique tracks and 2,002,482 records in total. We limited the Spotify dataset spanning the same period as the YouTube dataset: from August 12, 2020, to Februrary 28, 2023, except for South Korea. As Spotify launched their service in South Korea on February 1, 2021, we use the daily top chart data only after February 1, 2021, for South Korea.

For Netflix, we used a list of the top 10 most popular films and TV shows on Netflix (\url{https://top10.netflix.com}) retrieved on November 8, 2023. The dataset is dated every week from July 4, 2021, to February 28, 2023, including 18,100 TV shows and films in total. There are 920 unique TV shows and 1,940 unique movies. Note that NetFlix's charts are weekly charts, and thus the time resolution differs from the other datasets, yet one can compare the results because we aim to evaluate the long-term trends of the 600-days-long datasets rather than daily fluctuation (see Fig.~\ref{fig:sm} for the robustness of results regardless of the time). 

\subsection{Measuring socio-cultural distance between countries}
To examine the distance (or similarity) between countries, we employ additional socio-cultural datasets: i) Facebook social connectedness index (SCI)~\cite{bailey2021international} and ii) language lexicon similarity dataset~\cite{gabor2021lexicon} (see Fig.~\ref{fig:figs6} for the similarity between the two indices).

SCI data provides a normalized frequency of friendships between two countries on Facebook, which directly measures the degree of online social connection between countries. On the other hand, the language lexicon similarity dataset measures the similarity between two given languages using lexicon, which measures the distance (or barrier) between users of two languages regarding vocabulary. To project the language lexicon similarity at the country level, we also collected official language data from CIA World Factbook~\cite{cia-factbook} because the language lexicon similarity dataset does not give information about countries' spoken languages. To quantify the linguistic similarity between the two countries, we calculated the similarity ($LLS$) between country $i$ and country $j$ as follows:

\begin{equation}
    LLS_{ij} = \Sigma_{s \in L_i} \Sigma_{t \in L_j} w_{si} w_{tj} S(s,t),
\end{equation}
where $L_i$ represents the language set in country $i$, $w_{si}$ represents the share of language $s$ used in country $i$, and $S(s,t)$ is the lexicon similarity between language $s$ and language $t$ from the language lexicon similarity dataset~\cite{gabor2021lexicon}.

\subsection{Best fit model distributions of life time-series of contents}
To model the event's lifetime distribution, we commonly use power law and exponential distributions when it is highly skewed. For example, the exponential distribution is a suitable model for the decay of radioactive materials~\cite{istratov1999exponential}, whereas power law decay is an appropriate model for the aftershocks of earthquakes~\cite{narteau2005onset}. Such distributions are characterized by heavy tails, which makes it challenging to fit a suitable distribution from empirical data~\cite{clauset2009power}. The primary motivation for estimating these distributions is that, while they frequently have a similar visual appearance, understanding the precise distribution enables us to predict the mechanism governing popularity. For example, the power law indicates that popularity is decided by rewards like positive associations, whereas the lognormal arises from the process of repeatedly multiplying separate random distributions~\cite{mitzenmacher2004brief}. Therefore, to find the best-fit model distribution for the survival time of items in the trending chart, we choose five models that are frequently used to fit the skewed distributions and fit the empirical data using the maximum likelihood estimation as follows~\cite{alstott2014powerlaw}:
\begin{itemize}
\item{Power law
\begin{equation}
p(x) = (\alpha - 1)x^{\alpha-1}_\mathrm{min} x^{-\alpha},
\end{equation}
}
\item{Power law with an exponential cut-off
\begin{equation}
p(x) = \frac{\lambda^{1-\alpha}}{\Gamma(1-\alpha, \lambda x_\mathrm{min})} x^{-\alpha}e^{-\lambda x},
\end{equation}
}
\item{Exponential
\begin{equation}
p(x) = \lambda e^{\lambda x_\mathrm{min}} e^{-\lambda x},
\end{equation}
}
\item{Stretched exponential
\begin{equation}
p(x) =  \beta \lambda e^{\lambda x_\mathrm{min}^{\beta}} x^{\beta -1} e^{-\lambda x^{\beta}},
\end{equation}
}
\item{Lognormal
\begin{equation}
\begin{aligned}
p(x) = & \sqrt{\frac{2}{\pi \sigma^2}}\left[\mathrm{erfc}\left(\frac{\mathrm{ln} x_\mathrm{min}-\mu}{\sqrt{2}\sigma}\right)\right] ^{-1} \times \frac{1}{x}\mathrm{exp}\left[-\frac{(\mathrm{ln}x - \mu)^2}{2\sigma^2} \right].
\end{aligned}
\end{equation}
}
\end{itemize}

\section{Results}
\subsection{Co-trending contents between countries}

\begin{figure}
    \centering
    \includegraphics[width=0.5\textwidth]{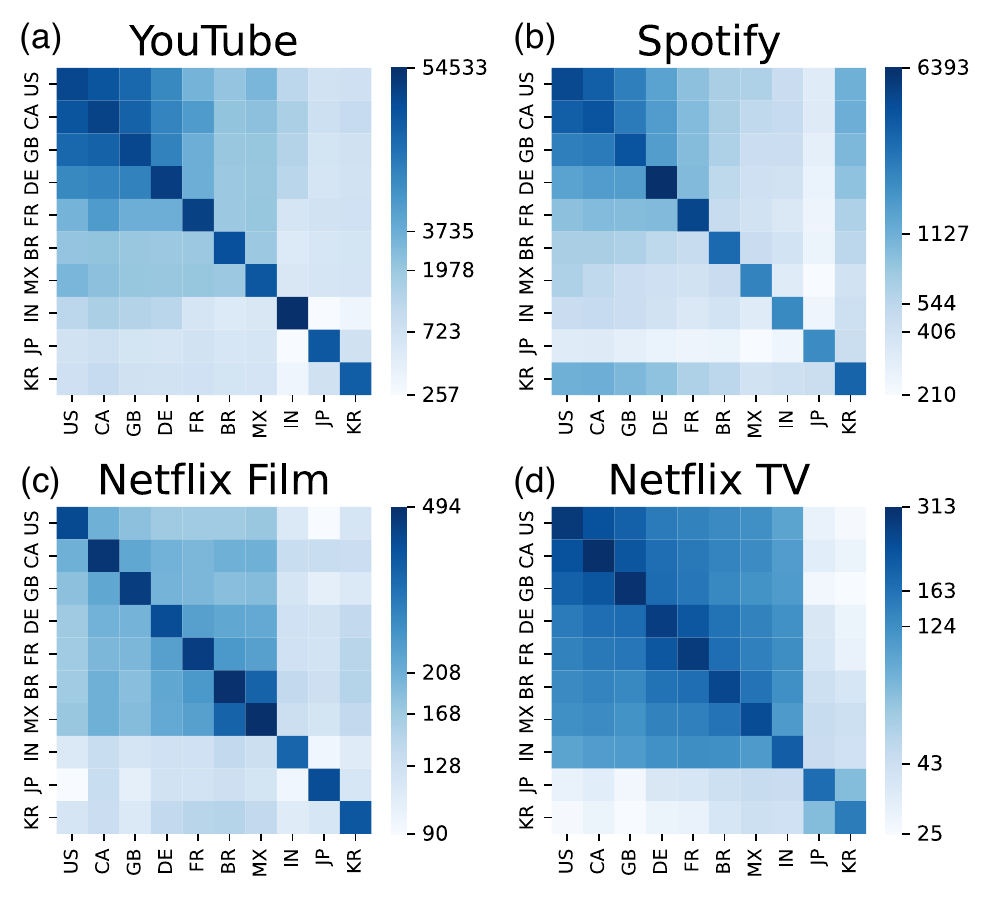}
    \caption{Numbers of shared trending content between countries: (a) YouTube, (b) Spotify, (c) Netflix Film, and (d) Netflix TV show. Each point is colored according to the number of shared contents in a log scale (see the color bar). The labels of the color bar correspond to the quartiles for each platform.}
    \label{fig:fig1}
\end{figure}

\begin{figure*}[htp!]
    \centering
    \includegraphics[width=\textwidth]{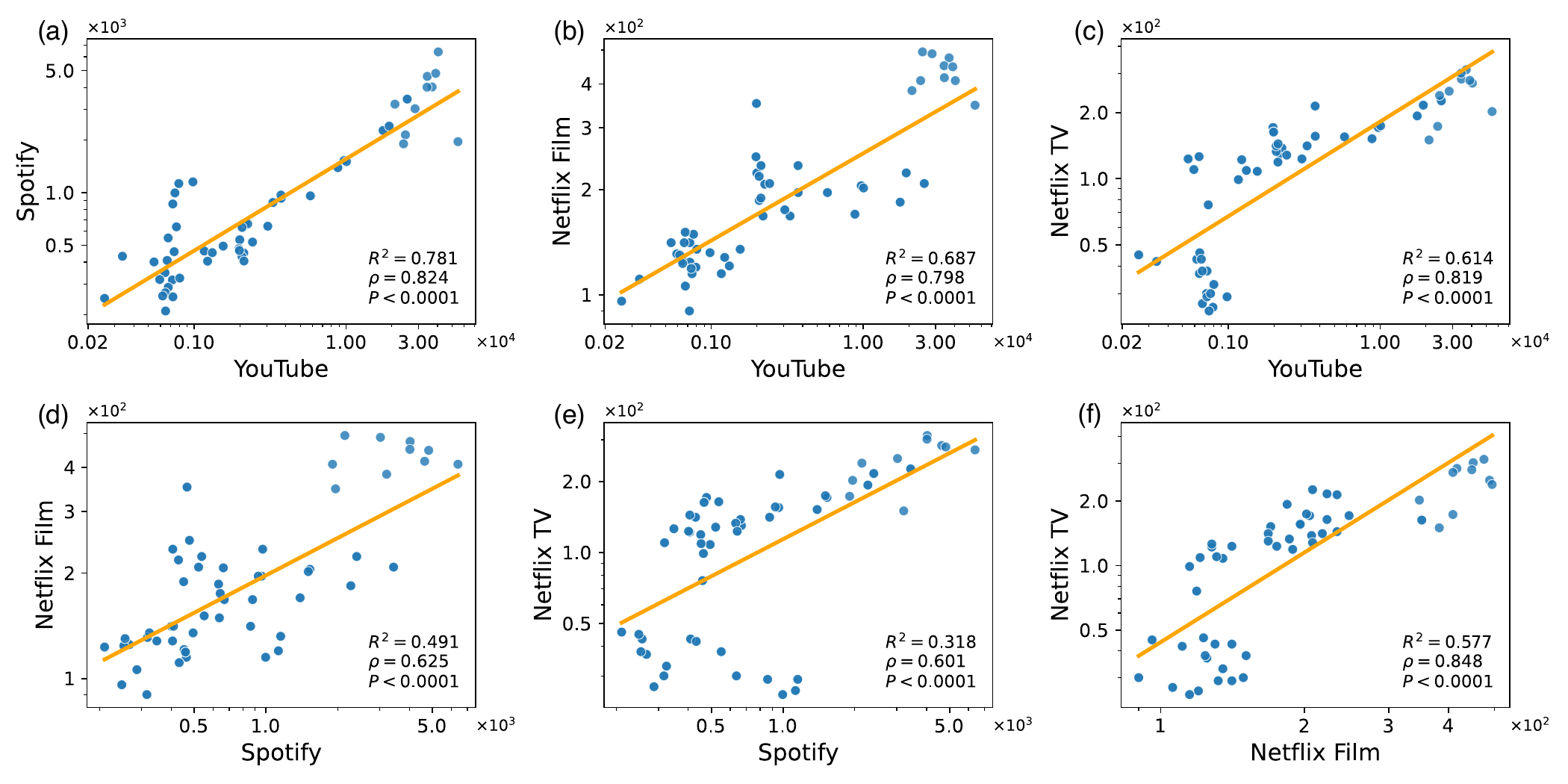}
    \caption{Cross-platform comparison for the numbers of co-trending contents between countries. The orange solid line represents a linear regression line between two platforms measured in a log-log scale ($y \sim x^k$), where we also measure the coefficient of determination ($R^2$) and Spearman correlation ($\rho$). The annotated p-value represents statistical significance tests for both $R^2$ and $\rho$. (a)--(c) While YouTube shows a high $R^2$ ($>0.6$) with every other platform, (d)--(e) Spotify and Netflix have a relatively lower coefficient of determination between them. (f) It's interesting to note that the relationship between Netflix's Film and the TV show has a lower $R^2$ than their relationship with YouTube. We observe similar patterns when we take into account the Spearman rank correlation.}
    \label{fig:fig2}
\end{figure*}

We first compared how many contents are consumed together between the two countries. Overall, most of trending contents is regional, which is consumed only in a single country. In Figure~\ref{fig:fig1}, the diagonal cells display the total number of trending contents for each country, and for most countries, the number of shared contents is not as many as the diagonal cells. We find that the United States, Canada, and the United Kingdom share a large number of contents on YouTube, Spotify, and Netflix TV (Figures~\ref{fig:fig1}(a), (b), and (d), respectively), while a lower number of content was trended together in Netflix Film (Figure~\ref{fig:fig1}(c)). Interestingly, on YouTube, Canada and France shared 5,821 contents, and the United States and Mexico shared 3,451 contents, which is a relatively large fraction of shared content compared to other countries' pairs. One possible scenario of this observation is the shared language user group between two countries. For instance, French is the official language in Quebec and Spanish is the second most spoken language in the United States.

One noteworthy point is that there seems a weaker language effect on the music streaming service (Spotify). The relative size of co-trending contents between Canada and France decreases, even though the number of co-trending contents between the United States and Mexico is large (Figure~\ref{fig:fig1}(b)). Instead, we also observe that the number of co-trended content between South Korea and other countries is relatively high. One may assume that this is mainly due to the rise of K-pop~\cite{choi2014k}, yet the large fraction, almost half, of the trending contents in South Korea are Western pop songs. This observation implies that trending charts in South Korea may be biased because of the user pool. Indeed, Spotify is not a major music streaming service in South Korea compared with local services, and Korean Spotify users may prefer Spotify to enjoy the wider Western pop song coverage.

When we move our attention to Netflix film (Figure~\ref{fig:fig1}(c)), there is a large block between Western countries. We also observe a distinguished number of co-trending contents between Mexico and Brazil. Asian countries present no significant tendency of co-trending between them. In contrast, the largest block in Netflix TV (Figure~\ref{fig:fig1}(d)) is composed of Western countries in addition to India. Using the hierarchical clustering, this largest block also can be divided into three groups: i) the United States, Canada, and the United Kingdom, ii) Germany and France, and iii) India, Brazil, and Mexico (See Supplementary Figure~\ref{fig:figs1}). Japan and South Korea co-consume a relatively small amount of contents, although they are geographically nearby; thus, in contrast to the previous study~\cite{brodersen2012youtube}, there are more influential factors that determine the spreading of streaming content.

The above results suggest that these possible factors that promote or inhibit content spreading are i) social connections or ii) language. Our observation also suggests that this co-trending tendency is different between streaming services, which provide different types of content from different producers. While Netflix mainly serves movies and TV series produced by professional video producers, Spotify serves musical recordings from professional musicians, curated by the service providers. On the other hand, anyone worldwide can share user-generated videos on YouTube. Therefore, from now on, we aim to determine the more influential factor between social connectedness and linguistic similarity in the spreading of streaming content by platforms and content types.

\subsection{Dynamics of trending contents by the platforms}

\begin{table*}[]
\centering
\caption{The best-fit distributions and their log-likelihoods for the total survival time distribution in the trending charts for each country (see Methods). The full fitting results of all five models can be found in Supplementary Tables~\ref{table:s1}--\ref{table:s4}}\label{table:fitting}.
\begin{tabular}{@{}ccccccccc@{}}
\toprule
         & \multicolumn{2}{c}{YouTube} 
         & \multicolumn{2}{c}{Spotify} 
         & \multicolumn{2}{c}{Netflix TV} 
         & \multicolumn{2}{c}{Netflix Film} \\ 
\makecell{Country\\(ISO2)}  & \makecell{Best\\distribution} & Loglikelihood 
& \makecell{Best\\distribution} & Loglikelihood 
& \makecell{Best\\distribution} & Loglikelihood 
& \makecell{Best\\distribution} & Loglikelihood \\
\midrule
BR  & SE & -63804.46  & TP & -13952.09 & LN    & -976.04  & LN    & -1506.11   \\
CA  & SE & -65111.92  & TP & -16828.35 & LN    & -1172.66  & LN    & -1486.51 \\
DE  & SE & -73416.77  & TP & -23780.35 & LN    & -1070.51  & LN    & -1377.75 \\
FR  & SE & -69299.91  & SE & -19838.77 & LN    & -1090.58  & LN    & -1444.85 \\
GB  & SE & -64322.76  & TP & -16224.22 & LN    & -1145.87  & LN    & -1457.60 \\
IN  & SE & -101575.88 & SE & -9858.39 & LN    & -857.64  & LN    & -1251.91  \\
JP  & SE & -52606.48  & SE & -9806.48 & LN    & -760.12  & LN    & -1386.82  \\
KR  & SE & -54897.41  & TP & -13807.96 & LN    & -697.07  & LN    & -1325.81 \\
MX  & SE & -55512.13  & TP & -10346.51 & LN    & -967.65  & LN    & -1516.86 \\
US  & LN & -67029.40  & TP & -18026.60 & LN    & -1093.63  & LN    & -1394.56 \\
\bottomrule
\end{tabular}

\vspace{1ex}
{\raggedright SE: Stretched exponential, TP: Truncated powerlaw, LN: Lognormal \par}
\end{table*}

To examine the dynamics of global streaming services, we first test the inertia of trending contents by calculating the total survival time in the trending chart. We then find the best-fit distribution of the total survival time distributions for each country by the platforms (see Methods), which are characterized by the size of the long tails. In the case of YouTube, all countries, except the United States, follow a stretched exponential distribution. The United States showed a lognormal distribution. For Spotify, different distributions were observed by countries. Brazil, Canada, Germany, Mexico, South Korea, the United Kingdom, and the United States demonstrate a truncated power law distribution; while France, India, and Japan showed a stretched exponential. Both Netflix films and TV shows presented a lognormal distribution for all ten countries. 

The findings suggest that user dynamics of contents follow different dynamics. Content following a truncated power law survival time distribution exhibits that contents can survive much longer when listed on the top charts but with natural decay because of the preferential attachment process. Therefore, users in these countries would be more sensitive to popular content. Content following a lognormal survival time distribution shows that content survival is a multiplicative process with independence. In other words, the content selection process is not governed by popularity, therefore the users would choose the content from their cultural preferences rather than popularity. In short, we can assume that content consumption can be affected by the socio-cultural similarity.

As a next step, we compare the content consumption similarity between streaming services (Figure~\ref{fig:fig2}). We computed the number of shared contents between the two countries and compared the correlation between the two streaming services. A simple linear regression between the two platforms shows that Spotify and YouTube exhibit the highest correlation ($R^2 = 0.781$). YouTube shows high correlations with all other platforms ($R^2=0.687$ and $0.614$ with Netflix Film and TV, respectively). Contrary to this, the correlation between Spotify and Netflix TV is relatively low ($R^2 = 0.318$, Figure~\ref{fig:fig2}(e)). One may consider the fact that Spotify serves audio-oriented content, whereas Netflix serves mainly video-oriented content. Human perceptions of audio-oriented and video-oriented content are different; so cultural proximity may have different influences on disseminating the content. It partially explains the low correlation between Spotify and Netflix TV. Going a step further, YouTube's strong relationships with other streaming services could be attributed to the platform's availability of audio-video hybrid content.

\subsection{Social Links vs. Language Barriers}

\begin{figure*}
    \centering
    \includegraphics[width=\textwidth]{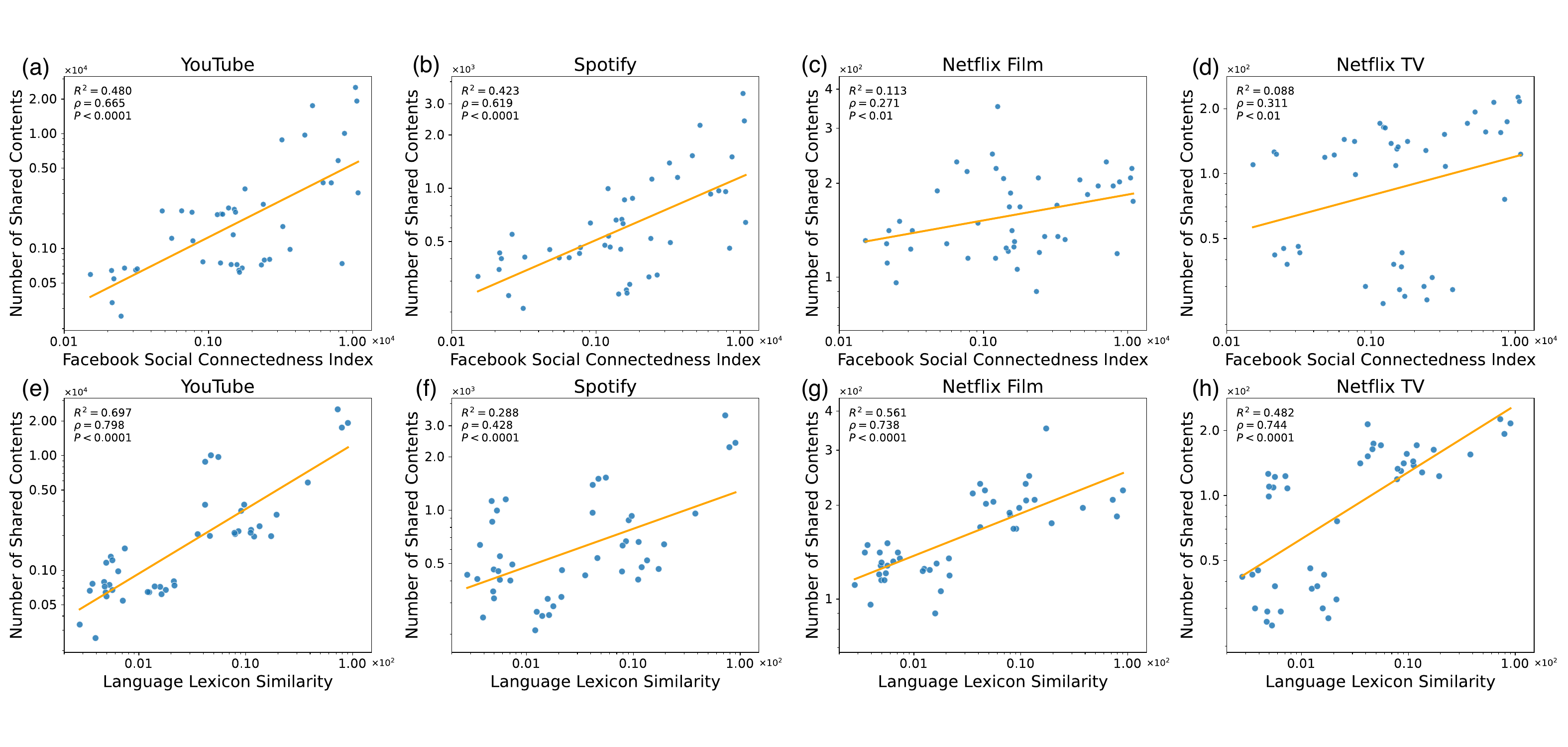}
    \caption{The correlation between the number of shared trending contents and two proxies of social similarity: (a)--(d) Facebook Social Connected Index~\cite{bailey2018social} and (e)--(h) language similarity (see Methods). The orange solid line represents a linear regression line between two platforms measured in a log-log scale ($y\sim x^k$), where we also measure the coefficient of determination ($R^2$) and Spearman correlation ($\rho$). The annotated p-value represents statistical significance tests for both $R^2$ and $\rho$. Spotify shows a comparatively stronger $R^2$ for social networks (Facebook SCI) than linguistic similarity (compare (b) with (f)), yet linguistic similarity displayed a greater $R^2$ for Netflix (compare (c)--(d) with (g)--(h)). YouTube shows high $R^2$ for both proxies of social similarity ((a) and (e))}
    \label{fig:fig3}
\end{figure*}

We extend the study to answer which socio-cultural factors play a more important role in trending content spreading. First, we apply the social connection as a proxy of social similarity~\cite{abisheva2014watches}. We used the Social Connectedness Index (SCI) provided by Meta, which describes the strength of the social connection between countries. When the SCI is high, the users in the two countries tend to be in a friendship. First, YouTube and Spotify show high correlations between SCI and the number of co-trending content (Figures~\ref{fig:fig3} (a) and (b), $R^2=0.480$ and $R^2=0.423$, respectively), whereas correlations between SCI and Netflix (both film and TV show) are less significant (Figures~\ref{fig:fig3} (c) and (d), $R^2=0.113$ and $R^2=0.088$, respectively). In short, when the two countries are socially connected, they tend to share the same audio-oriented trending content more.

While the social tie is correlated with the spreading of audio-oriented content, we also find that linguistic similarity influences more to the spreading of video-oriented content. YouTube, Netflix Film, and Netflix TV shows are highly correlated with language lexicon similarity ($R^2 = 0.697$, $0.561$, and $0.482$ in Figures~\ref{fig:fig3} (e), (g), and (h), respectively). One possibility is that to fully understand video-oriented content, such as films and TV shows, one requires the ability to understand nuanced expressions so that users in the same language can be more familiar with the content. However, because of the lack the user information, we left it for further study. Contrary to this, Spotify exhibits a lower correlation than others ($R^2 = 0.288$, Figure~\ref{fig:fig3}(f)). This result implies that the spreading of audio-oriented content, such as music, is influenced less by the language than that of video-oriented content.

Combining the results, we can summarize that the influential factor of content spreading depends on the type of content. The spreading of auditory content is highly correlated with social connection, whereas the spreading of visual content is limited by language. One interesting point is that YouTube shows a dual characteristic of audio-based and video-based content. Indeed, YouTube has a various type of content uploaded by individuals. In short, YouTube's high correlation with both social connections and linguistic similarity may come from YouTube's wide range of content types; however, it also can be due to the unique characteristics of YouTube. Netflix and Spotify are a sort of alternative service to legacy media. Netflix serves as an alternative for theaters and Spotify serves as a substitute for music media such as compact discs. However, YouTube does not have such a counterpart. Therefore it necessitates a more in-depth analysis of two possible reasons. 

\subsection{Decomposing YouTube into the topical categories}\label{subsec:categories}

\begin{figure*}
    \centering
    \includegraphics[width = \textwidth]{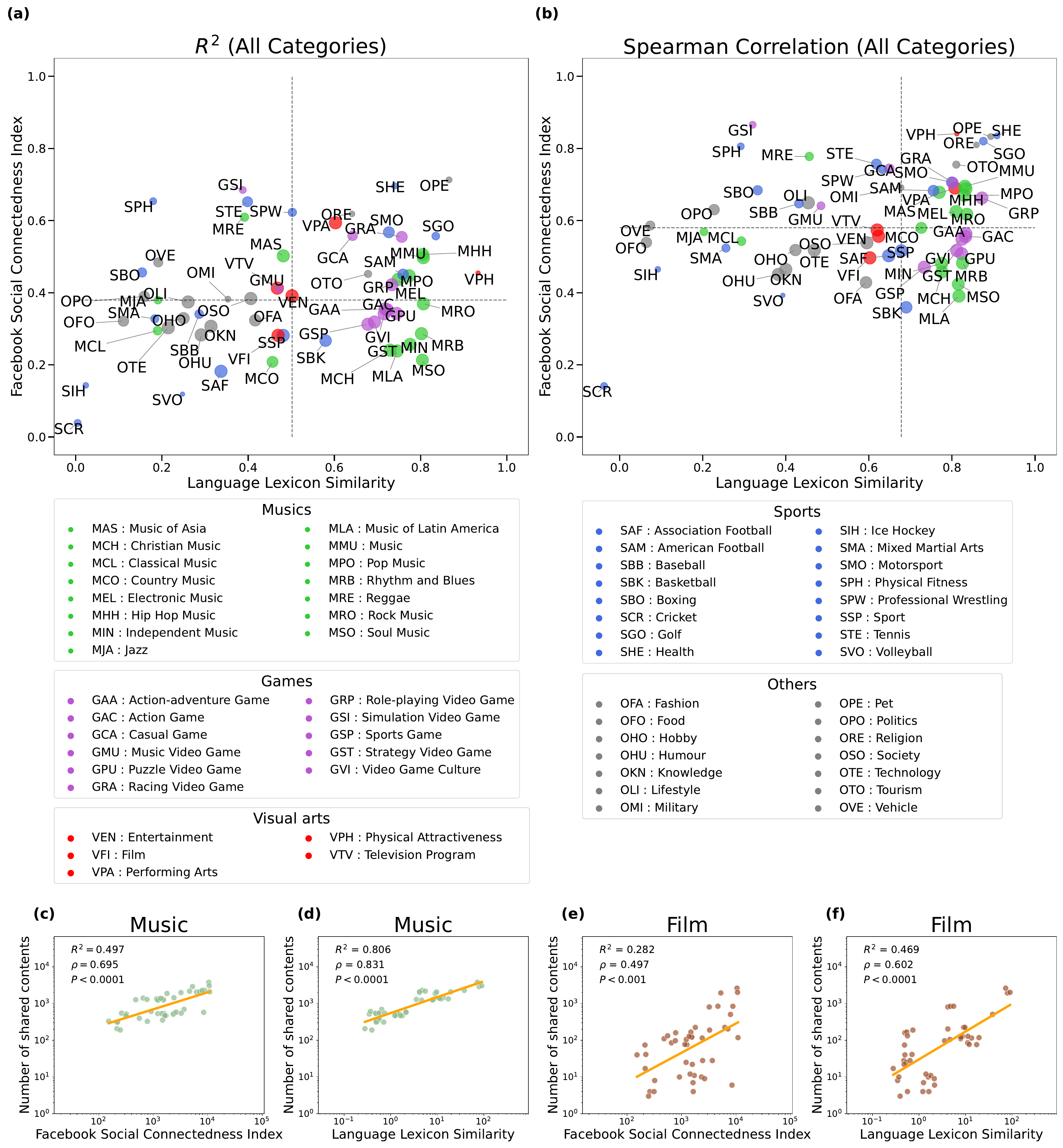}
    \caption{Interrelationship between the correlation of the trending videos on YouTube with the Facebook SCI and the language similarities across categories. For (a), the x-axis represents the $R^2$ of the number of shared trending videos to the Facebook SCI between countries, whereas the y-axis represents the $R^2$ between the number of shared trending videos to the language similarity. Panel (b) shows similar relations using the Spearman rank correlation instead of $R^2$. For panels (a) and (b), the color of circles represents the category groups: Musics categories (green), Games categories (purple), Sports categories (blue), Visual arts (red), and others (grey). The abbreviation and full name of each category are described in the legend. See Supplementary Table~\ref{tab:tab1} for the full list of categories and Tables~\ref{table:s6} and ~\ref{table:s7} for the detailed statistics. The diameter of the circles corresponds to the number of country pairs that have mutually shared trending videos. The dashed line in (a) and (b) represents a median value of language lexicon similarity and SCI of all categories in the data. Panels (c)--(f) display the scatter of the number of shared trending videos with Facebook SCI ((c) and (e)) and language similarity ((d) and (f)), respectively, for two example categories: (c)--(d) \texttt{Music} and (e)--(f) \texttt{Film}. }
    \label{fig:fig4}
\end{figure*}

We step into analyzing categories in YouTube content to figure out the underlying reason behind the high correlation of YouTube for both social connections and linguistic similarity. To do this, we assign the categories for each content using YouTube Data API (See Methods). Figures~\ref{fig:fig4} (a) and (b) display $R^2$ and Spearman rank correlation of the number of co-trending videos with SCI and language lexicon similarity between countries, by the category. We found that the correlation varies by the categories. For instance, contents in the musical category have a high correlation with language lexicon similarity (green circles in Figures~\ref{fig:fig4}(a) and (b)), while contents in the sports category have a large variance with both language lexicon similarity and SCI (blue circles).

In the musics category group, YouTube contents show a higher correlation with language similarity than the SCI, where the median $R^2$ is $0.748$ and $0.369$ for language lexicon similarity and SCI, respectively. Recall the observation that Spotify shows a higher correlation with the SCI than language lexicon similarity (Figure~\ref{fig:fig3}), YouTube's musics category groups displays different characteristics to Spotify. If one looks in detail, one may find that \texttt{Reggae}, \texttt{Jazz}, and \texttt{Classical music} show low $R^2$ with language lexicon similarity, which is consistent with Spotify's results, yet many other musical categories show a high correlation with language lexicon similarity.

We also find that most of the games category groups exhibit a high correlation with the language lexicon similarity (median $R^2 = 0.716$). Since the gaming contents are visual-centered with the narrative and textual elements within games (such as dialogue and storylines), linguistic factors play a crucial role in their global dissemination. There is no noticeable influence of social connections or linguistic similarity for the sports category, although our dataset includes important sports events such as the \textit{Tokyo 2020 Summer Olympics}, \textit{FIFA World Cup Qatar 2022}, and \textit{Beijing 2022 Winter Olympics}. One potential cause is that each country has distinct broadcasting companies that own regionally exclusive broadcasting rights; there can be identical events yet from different sources.

One interesting point is that the correlation with the linguistic similarity of the visual arts category group (median $R^2 = 0.502$) is lower than that of the musics category group (median $R^2 = 0.748$), although it shows similar correlations with Netflix film and TV show ($R^2 = 0.561$ and $0.482$, respectively; see Figure~\ref{fig:fig3}). In addition, the correlation with the SCI of the visual arts category group (median $R^2 = 0.413$) is high, compared to Netflix films ($R^2 = 0.113$) and TV shows ($R^2 = 0.088$). Indeed, we can see the clear correlations in Figures~\ref{fig:fig4}(c)--(f) compared to Figures~\ref{fig:fig3}(b)--(c) and (f)--(g). These findings also support our former finding that YouTube, as a completely new type of media, shows different characteristics compared with the alternatives of legacy media; YouTube shows a dual characteristic of visual-oriented and audio-oriented content.

\section{Discussion}

What streaming platforms spread is not only in the media contents but also the cultures, which facilitate communication between communities with different \textit{habitus} in contemporary society. Previous studies mainly focused on individual platforms or within specific countries~\cite{pinto2013using}. Although there are attempts to analyze the global perspective~\cite{abisheva2014watches}, it mainly considers the impact of geographical barriers~\cite{brodersen2012youtube}, which is now gradually diminished~\cite{yoon2023quantifying}. Our study takes a wider angle of view by analyzing co-trending content between ten countries in the three most popular global streaming services: YouTube, Spotify, and Netflix. We then try to elucidate the underlying influential factors of spreading content by using social connection and linguistic affinity.

Our findings suggest that intercountry content spreading patterns are different by streaming services. To elucidate the underline factors influencing the spreading, we employed two proxies of social similarity, SCI representing the tendency of direct friendship between countries' populations and lexicon similarity accounting for the similarity and barrier due to the languages~\cite{yoon2023quantifying}. The spreading of music (or auditory content), evidenced by Spotify, is largely influenced by social connectivity and insignificantly influenced by linguistic barriers. Dissemination of video content, observed from Netflix, depends more on language rather than social connectivity. Language and social connectivity both show a large influence on content spreading on YouTube. One may suppose that this is because YouTube contains both auditory and visual content, yet our study shows that the platform's strong correlation is not due to this because language and social connectivity have a significant impact on the spreading of both music-focused and visual-focused categories on YouTube. Instead, due to the unique prosumer behavior in YouTube, which one being both producer and consumer simultaneously~\cite{holland2016youtube}, users may be more tightly engaged in social connection, while the language similarity facilitates the spreading of the contents. Therefore, YouTube establishes a unique ecosystem, rather than a mixture of legacy ecosystem of music and video content separating the consumers and producers. 

We believe that such data have considerable potential for future research also. In this study, we use trending content in streaming platforms, which covers a relatively small number of content concerning the entire volume of content in the platforms; in addition, although we selected three well-known, and global, streaming platforms there are thousands of other streaming platforms, and thus we hardly cover entire user pool. In addition, different recommendation algorithms and content curation in each platform can affect the content trending. Their recommendation algorithm suggests the contents that may fit the regional characteristics. If users depend on the recommended content, the trending content is just the result of recommendations. Since Netflix provides different content for each country, due to the copyright, content curation may limit the number of shared content. Our findings are possibly due to the limited user pools. For instance, because Spotify is not a major service in South Korea, the user pool in South Korea is biased toward heavy listeners preferring Western music. Additionally, the differences can be from the business model, that is paid-subscription model for Spotify and Netflix compared with the advertising-subscription hybrid model of YouTube. Because we only use the degree of friendship as the proxy of social connection, the implication of the study will enhance with the additional analysis on the actual spreading behavior ~\cite{berger2012makes, weng2013virality}, along with the detailed analysis on the socio-cultural background of the group of users based on their platform selection and living country.

We demonstrate how language usage and social connections affect the spreading of online content, suggesting that individuals may react differently to the same content based on their backgrounds. Thus, quantifying the differences in interest changes based on their social background and language may be beneficial to understanding the hidden pattern of human behavior. In addition, distinct patterns of dissemination between user-generated videos and conventional content highlight the content consumption dependence on new features of content. We expect that emerging technologies such as AR and VR have a potential to introduce a new content consumption pattern.Content producers and platform providers can also benefit from our research in a practical manner. The cost of content translation and review forces the provider to choose content that generates more views for the same amount of money, and it is possible if we consider the kind of content and its social and cultural ties. Predicting the virality of individual contents necessitates more investigation, which we leave for future study. By spotlighting the influential factors of cultural spreading, we want to shed light on the unexplored mechanism underlying the general rules of cultural spreading and adoption. Finally, we emphasize that our research has potential wider implications in contemporary society, not restricted to streaming platforms, as soft power is increasingly important in contemporary society~\cite{nye1990soft}. 

\bibliographystyle{ACM-Reference-Format}
\bibliography{cite}

\begin{acks}

This work was supported by the National Research Foundation of Korea (NRF) funded by the Korean government (grant No. NRF-2022R1C1C2004277 (T.Y.) and 2022R1A2C1091324 (J.Y.)). The Korea Institute of Science and Technology Information (KISTI) also supported this research by providing KREONET, a high-speed Internet connection. This work was also supported by Innovative Human Resource Development for Local Intellectualization program through the Institute of Information \& Communications Technology Planning \& Evaluation(IITP) grant funded by the Korea government(MSIT) (IITP-2024-RS-2022-00156360). The funders had no role in study design, data collection and analysis, decision to publish, or preparation of the manuscript. Google Bard (\url{https://bard.google.com}) was used to rephrase the title and Grammarly (\url{https://grammarly.com/}) was used to check the grammatical error of the manuscript. The authors also have declared that no competing interests.

\end{acks}

\pagebreak

\onecolumn 

\setcounter{page}{1}
\setcounter{figure}{0}
\setcounter{table}{0}
\setcounter{section}{0}

\renewcommand{\thesection}{S\arabic{section}}
\renewcommand{\thetable}{S\arabic{table}}
\renewcommand{\thefigure}{S\arabic{figure}}

\centering
\noindent\huge{\textbf{Supplementary Material\\``Social Links vs. Language Barriers:\\Decoding the Global Spread of Streaming Content''}}

\vspace{0.3cm}

\Large{Seoyoung Park, Sanghyeok Park, Taekho You, Jinhyuk Yun}

\section{Supplementary Tables}\label{section:SITables}

\begin{table}[ht]
\centering
\caption{The full fitting results (log-likelihoods) for the total survival time distribution in YouTube's trending chart for each country. Boldfaced values in blue indicate the best (maximum), and values in red indicate the second best log-likelihood values for each country. Ratio is the result of loglikelihood ratio test between the best and the second best distribution using powerlaw package.}\label{table:s1}
\begin{threeparttable}
\begin{tabular}{@{}ccccccc@{}}
\toprule
Country (ISO2) & Power law  & Lognormal  & Exponential & \makecell{Truncated\\Power Law} & \makecell{Stretched\\Exponential} & Ratio \\ \midrule
BR             & -102891.46 & \textbfg{-68164.69}          & -79029.07   & -91077.16           & \textbfm{-63804.46}  & 53.8843 ***  \\
CA             & -121425.26 & \textbfg{-67289.71}          & -91861.04   & -116098.97          & \textbfm{-65111.92}  & 16.0750 ***  \\
DE             & -126921.94 & \textbfg{-75184.64}          & -96731.40   & -111368.92          & \textbfm{-73416.76}  & 16.5116 ***  \\
FR             & -124663.94 & \textbfg{-72094.28}          & -94703.39   & -109396.14          & \textbfm{-69299.91}  & 27.8654 *** \\
GB             & -115820.08 & \textbfg{-69401.40}          & -88059.68   & -101942.40          & \textbfm{-64322.76}  & 45.1209 ***  \\
IN             & -133461.96 & \textbfg{-103769.69}         & -108682.50  & -118440.42          & \textbfm{-101575.88}  &  49.8608 *** \\
JP             & -93623.47  & \textbfg{-55808.74}          & -70993.65   & -82915.97           & \textbfm{-52606.48}  &  30.6246 ***  \\
KR             & -84070.91  & \textbfg{-57658.57}          & -64803.40   & -74827.12           & \textbfm{-54897.41}  &  37.1456 ***  \\
MX             & -94762.00  & \textbfg{-59287.60}          & -72164.98   & -83945.82           & \textbfm{-55512.13} & 34.1489 ***  \\
US             & -116215.82 & \textbfm{-67029.40} & -88201.93   & -111579.68          & \textbfg{-72856.86}  &  8.2596 ***  \\ \bottomrule
\end{tabular}
\begin{tablenotes}
    \item \textbf{Note} * ** *** indicate levels of significance at 0.90, 0.95, 0.99, respectively.
\end{tablenotes}
\end{threeparttable}
\end{table}

\pagebreak

\begin{table}[ht]
\centering
\caption{The full fitting results (log-likelihoods) for the total survival time distribution in Spotify's trending chart for each country. Boldfaced values in blue indicate the best (maximum), and values in red indicate the second best log-likelihood values for each country. Ratio is the result of loglikelihood ratio test between the best and the second best distribution using powerlaw package.}\label{table:s2}
\begin{threeparttable}
\begin{tabular}{@{}cccccccc@{}}
\toprule
Country (ISO2) & \multicolumn{1}{c}{Power law} & \multicolumn{1}{c}{Lognormal} & \multicolumn{1}{c}{Exponential} & \multicolumn{1}{c}{\makecell{Truncated\\Power Law}} & \multicolumn{1}{c}{\makecell{Stretched\\Exponential}} & \multicolumn{1}{c}{Ratio}\\ \midrule
BR             & -14639.83                     & -14145.15                     & -15449.11                       & \textbfm{-13952.09}                      & \textbfg{-14028.58}          & 9.6292 ***                       \\
CA             & -17436.48                     & -16905.47                     & -19418.04                       & \textbfm{-16828.35}                      & \textbfg{-16853.03}          & 3.9347 ***                       \\
DE             & -24482.26                     & -23877.51                     & -27836.62                       & \textbfm{-23780.35}                      & \textbfg{-23806.93}          & 3.5766 ***                       \\
FR             & -20667.34                     & -19894.28                     & -22358.69                       & \textbfg{-19840.52}                      & \textbfm{-19838.77}          & 0.2326 \tnote{1}             \\
GB             & -16681.96                     & -16347.36                     & -19383.72                       & \textbfm{-16224.22}                      & \textbfg{-16286.18}          & 11.5625 ***                       \\
IN             & -10394.80                     & -9934.51                      & -10756.38                       & \textbfg{-9862.80}                       & \textbfm{-9858.39}           & 0.4164 \tnote{2}                       \\
JP             & -10391.18                     & -9876.22                      & -10610.86                       & \textbfg{-9838.63}                       & \textbfm{-9806.48}           & 2.6256 ***                       \\
KR             & -14410.49                     & -13851.83                     & -15528.73                       & \textbfm{-13807.96}                      & \textbfg{-13816.69}          & 1.4089 \tnote{3}                       \\
MX             & -10815.51                     & -10482.88                     & -11693.73                       & \textbfm{-10346.51}                      & \textbfg{-10409.17}          & 11.5227 ***                       \\
US             & -18497.17                     & -18115.71                     & -21680.07                       & \textbfm{-18026.60}                      & \textbfg{-18070.63}          & 7.5631 ***                       \\ \bottomrule
\end{tabular}
\begin{tablenotes}
    \item \textbf{Note} * ** *** indicate levels of significance at 0.90, 0.95, 0.99, respectively.
    \item[1] P = 0.8161 \item[2] P = 0.6771 \item[3] P = 0.1588
\end{tablenotes}
\end{threeparttable}
\end{table}

\begin{table}[hb]
\centering
\caption{The full fitting results (log-likelihoods) for the total survival time distribution in Netflix's film trending chart for each country. Boldfaced values in blue indicate the best (maximum), and values in red indicate the second best log-likelihood values for each country. Ratio is the result of loglikelihood ratio test between the best and the second best distribution using powerlaw package.}\label{table:s3}
\begin{threeparttable}
\begin{tabular}{@{}ccccccc@{}}
\toprule
Country (ISO2) & Power law & Lognormal         & Exponential & \makecell{Truncated\\Power Law} & \makecell{Stretched\\Exponential} & Ratio          \\ \midrule
BR             & -2188.90  & \textbfm{-1506.11} & -1699.66    & -1960.98            & \textbfg{-1564.61}              & 10.9001 ***    \\
CA             & -2139.43  & \textbfm{-1486.51} & -1667.74    & -1918.58            & \textbfg{-1566.07}              & 7.3722 ***     \\
DE             & -1903.25  & \textbfm{-1377.75} & -1499.53    & -1712.84            & \textbfg{-1437.75}              & 7.0992 ***     \\
FR             & -2049.75  & \textbfm{-1444.85} & -1600.21    & -1839.94            & \textbfg{-1496.85}              & 4.8611 ***     \\
GB             & -2047.50  & \textbfm{-1457.60} & -1607.80    & -1839.55            & \textbfg{-1534.89}              & 7.3525 ***     \\
IN             & -1648.84  & \textbfm{-1251.91} & -1333.91    & -1492.79            & \textbfg{ -1322.67}              & 5.3124 ***     \\
JP             & -1892.31  & \textbfm{-1386.82} & -1499.53    & -1704.60            & \textbfg{-1451.61}              & 5.5812 ***     \\
KR             & -1807.39  & \textbfm{-1325.81} & -1430.21    & -1628.91            & \textbfg{-1381.24}              & 1.4329 \tnote{1}\\
MX             & -2197.07  & \textbfm{-1516.86} & -1714.26    & -1969.62            & \textbfg{-1623.70}              & 5.8921 ***     \\
US             & -1917.63  & \textbfm{-1394.56} & -1520.50    & -1727.21            & \textbfg{-1483.24}              & 8.0481 ***     \\ \bottomrule
\end{tabular}
\begin{tablenotes}
    \item \textbf{Note} * ** *** indicate levels of significance at 0.90, 0.95, 0.99, respectively.
    \item[1] P = 0.15
\end{tablenotes}
\end{threeparttable}
\end{table}
\pagebreak

\begin{table}[ht]
\centering
\caption{The full fitting results (log-likelihoods) for the total survival time distribution in Netflix's film trending chart for each country. Boldfaced values in blue indicate the best (maximum), and values in red indicate the second best log-likelihood values for each country. Ratio is the result of loglikelihood ratio test between the best and the second best distribution using powerlaw package.}\label{table:s4}
\begin{threeparttable}
\begin{tabular}{@{}ccccccc@{}}
\toprule
Country (ISO2) & Power law & Lognormal         & Exponential & \makecell{Truncated\\Power Law} & \makecell{Stretched\\Exponential} & Ratio      \\ \midrule
BR             & -2188.90  & \textbfm{-1506.11} & -1699.66    & -1960.98            & \textbfg{-1564.61}              & 8.5722 *** \\
CA             & -2139.43  & \textbfm{-1486.51} & -1667.74    & -1918.58            & \textbfg{-1566.07}              & 6.9314 *** \\
DE             & -1903.25  & \textbfm{-1377.75} & -1499.53    & -1712.84            & \textbfg{-1437.75}              & 8.1872 *** \\
FR             & -2049.75  & \textbfm{-1444.85} & -1600.21    & -1839.94            & \textbfg{-1496.85}              & 7.5028 *** \\
GB             & -2047.50  & \textbfm{-1457.60} & -1607.80    & -1839.55            & \textbfg{-1534.89}              & 11.1907 *** \\
IN             & -1648.84  & \textbfm{-1251.91} & -1333.91    & -1492.79            & \textbfg{-1322.67}              & 12.1889 *** \\
JP             & -1892.31  & \textbfm{-1386.82} & -1499.53    & -1704.60            & \textbfg{-1451.61}              & 12.3915 *** \\
KR             & -1807.39  & \textbfm{-1325.81} & -1430.21    & -1628.91            & \textbfg{-1381.24}              & 8.8918 *** \\
MX             & -2197.07  & \textbfm{-1516.86} & -1714.26    & -1969.62            & \textbfg{-1623.70}              & 10.9037 *** \\
US             & -1917.63  & \textbfm{-1394.56} & -1520.50    & -1727.21            & \textbfg{-1483.24}              & 5.7429 *** \\ \bottomrule
\end{tabular}
\begin{tablenotes}
    \item \textbf{Note} * ** *** indicate levels of significance at 0.90, 0.95, 0.99, respectively.
\end{tablenotes}
\end{threeparttable}
\end{table}

\begin{table}[hb]
  \centering
  \caption{Full category list in Figure~\ref{fig:fig4}}
  \label{table:s5}
  \begin{tabular}{cc}
    \toprule
    Category group&Categories\\
    \midrule
    \textbf{Musics}&\makecell[c]{
    Christian Music, Country Music, Classical Music, Electronic Music, Hip Hop Music,\\
    Independent Music, Jazz, Music, Music of Asia, Music of Latin America,\\
    Pop Music, Reggae, Rhythm and Blues, Rock Music, Soul Music (15)}\\
    \midrule
    \textbf{Games}&\makecell[c]{
    Action Game, Action-adventure Game, Casual Game, Music Video Game,\\
    Puzzle Video Game, Role-playing Video Game, Racing Video Game, Simulation Video Game,\\
    Sports Game, Strategy Video Game, Video Game Culture (11)}\\
    \midrule  
    \textbf{Sports}&\makecell[c]{
    American Football, Association Football, Baseball, Basketball, Boxing, Cricket, Golf, Health,\\ 
    Ice Hockey, Mixed Martial Arts, Motorsport, Professional Wrestling, Physical Fitness, \\
    Sport, Tennis, Volleyball (16)}\\
    \midrule
    \textbf{Visual arts}&\makecell[c]{
    Entertainment, Film, Television Program, Performing Arts, Physical Attractiveness (5)}\\
    \midrule
    \textbf{Others}&\makecell[c]{
    Fashion, Food, Hobby, Humour, Knowledge, Lifestyle, Military, Pet, Politics, Religion,\\
    Society, Technology, Tourism, Vehicle (14)}\\
  \bottomrule
\end{tabular}
\end{table}

\pagebreak

\begin{table}[H]
\centering
\caption{The median values of categories through $R^2$ for the types of category. "Number of Links" is the number of country pairs that have mutually shared trending videos. }\label{table:s6}
\begin{tabular}{cccccc}
\toprule
Category group &  \makecell[c]{Facebook SCI \\ $R^2$} & \makecell[c]{Facebook SCI \\ p-value} &  \makecell[c]{Language \\ Lexicon Similarity \\ $R^2$} & \makecell[c]{Language \\ Lexicon Similarity \\ p-value} &  Number of Links \\
\midrule
        \textbf{Musics} & 0.369165 *** & 0.000152 & 0.748075 *** & 1.861483e-14 & 45.0 \\
        \textbf{Games} & 0.355107 *** & 0.000024 & 0.715712 *** & 1.340331e-12 & 45.0 \\
        \textbf{Sports} & 0.395615 *** & 0.000508 & 0.367764 *** & 4.985299e-04 & 22.0 \\
        \textbf{Visual arts} & 0.412635 *** & 0.000004 & 0.501867 *** & 2.080118e-07 & 45.0 \\
        \textbf{Others} & 0.378200 *** & 0.000131 & 0.302363 *** & 7.652653e-04 & 36.0 \\
        \hdashline
        \textbf{All} & 0.379361 *** & 0.000134 & 0.501867 *** & 6.131285e-06 & 40.0 \\
\bottomrule
\end{tabular}
\begin{tablenotes}
    \item \textbf{Note} * ** *** indicate levels of significance at 0.90, 0.95, 0.99, respectively.
\end{tablenotes}
\end{table}

\begin{table}[H]
\centering
\caption{The median values of categories through  Spearman's correlation for the types of category. "Number of Links" is the number of counrty pairs that have mutually shared trendgin videos. (Same as \textbf{Table S6}.)}\label{table:s7}

\begin{tabular}{cccccc}
\toprule
Category group & \makecell[c]{Facebook SCI \\ Spearman's \\ Correlation} & \makecell[c]{Facebook SCI \\ p-value} & \makecell[c]{Language \\ Lexicon Similarity \\ Spearman's Correlation} & \makecell[c]{Language \\ Lexicon Similarity \\  p-value} &  Number of Links \\
\midrule
          \textbf{Musics} & 0.579752 *** & 0.000211 & 0.809790 *** & 2.824521e-11 & 45.0 \\
           \textbf{Games} & 0.563484 *** & 0.000074 & 0.811503 *** & 1.365777e-11 & 45.0 \\
          \textbf{Sports} & 0.665081 *** & 0.000361 & 0.625595 *** & 1.167764e-03 & 22.0 \\
         \textbf{Visual arts} & 0.573895 *** & 0.000072 & 0.623250 *** & 5.615586e-06 & 45.0 \\
          \textbf{Others} & 0.562797 *** & 0.001255 & 0.461339 *** & 4.120766e-03 & 36.0 \\
          \hdashline
          \textbf{All} & 0.579752 *** & 0.000287 & 0.677976 *** & 1.624084e-05 & 40.0 \\
\bottomrule
\end{tabular}
\begin{tablenotes}
    \item \textbf{Note} * ** *** indicate levels of significance at 0.90, 0.95, 0.99, respectively.
\end{tablenotes}
\end{table}

\newpage
\section{Supplementary Figures}\label{section:SIFigs}

\begin{figure}[H]
    \centering
    \includegraphics[width = 0.8\textwidth]{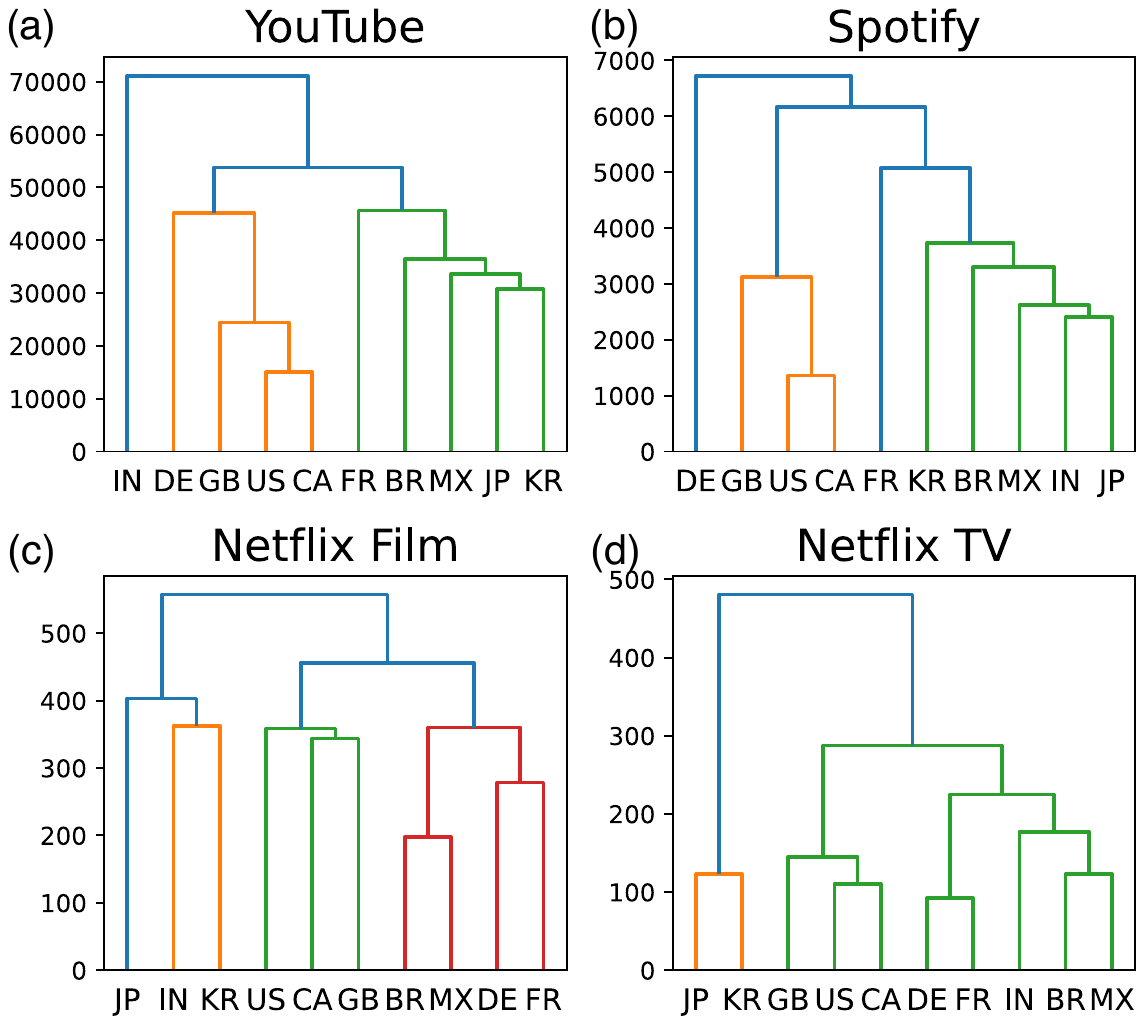}
    \caption{The dendrogram for the complete-linkage cluster of Figure~\ref{fig:fig1}: (a) YouTube, (b) Spotify, (c) Netflix Film, and (d) Netflix TV show. The cluster was optained using the Scipy's \texttt{scipy.cluster.hierarchy.linkage} with following parameters: \texttt{method='complete', metric='euclidean', optimal\_ordering=False})}
    \label{fig:figs1}
\end{figure}

\begin{figure}[H]
    \centering
    \includegraphics[width = 0.95\textwidth]{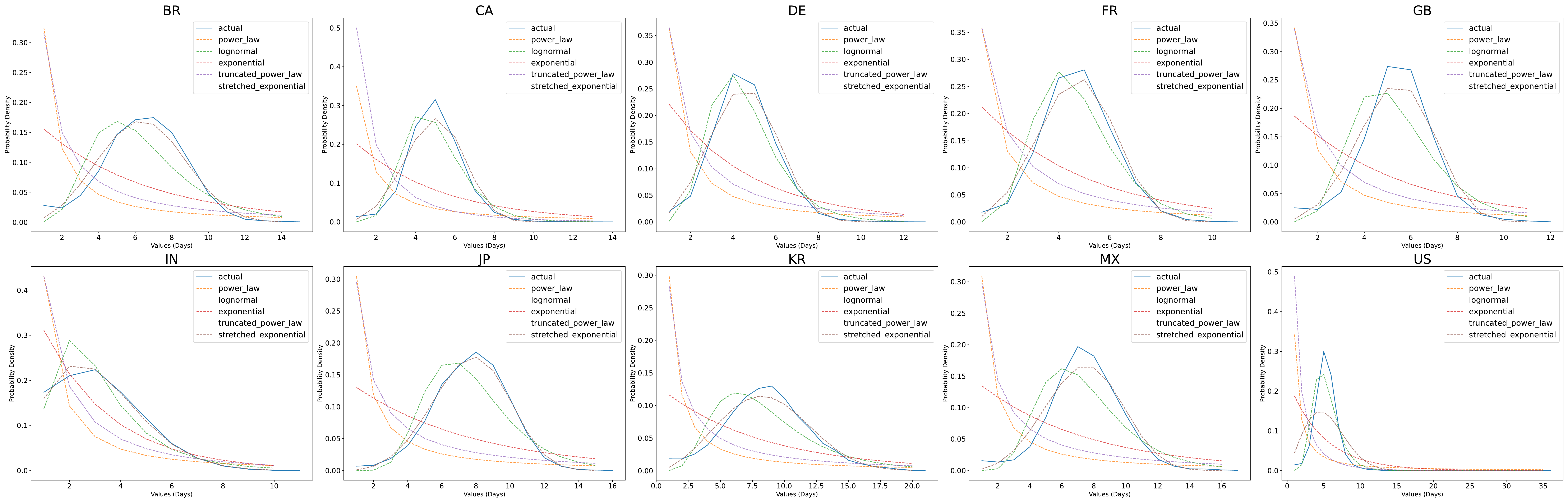}
    \caption{The fitted curve for the total survival time distribution in YouTube's trending chart for each country. We test five model distributions (see Methods).}
    \label{fig:figs2}
\end{figure}

\begin{figure}[H]
    \centering
    \includegraphics[width = 0.95\textwidth]{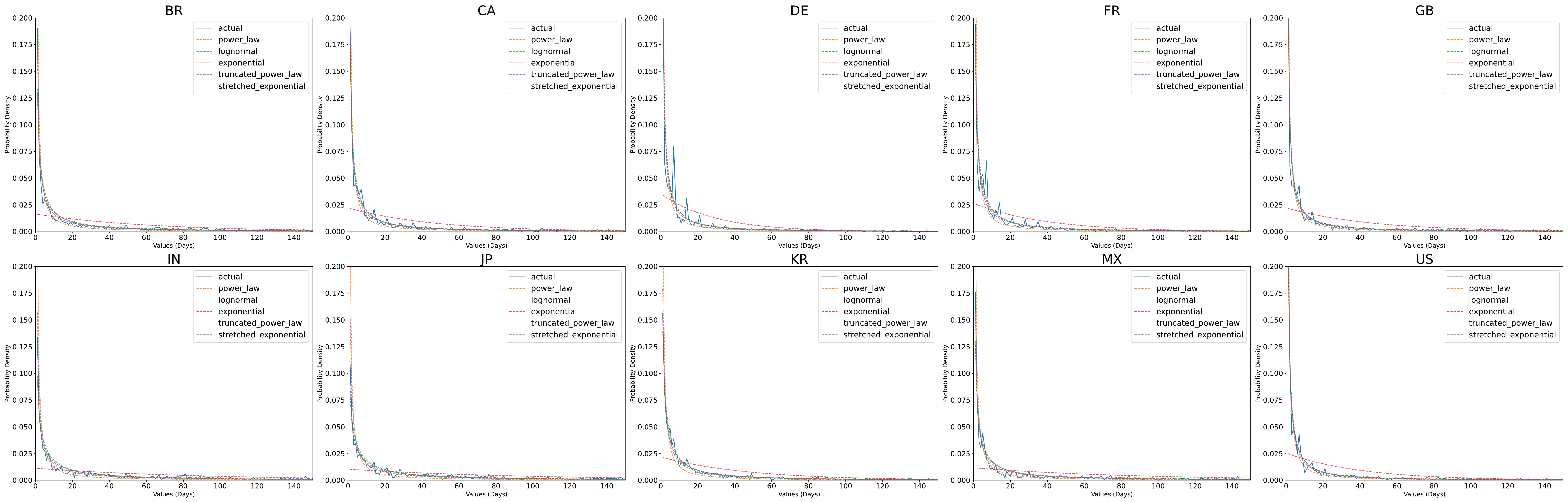}
    \caption{The fitted curve for the total survival time distribution in Spotify's trending chart for each country. We test five model distributions (see Methods).}
    \label{fig:figs3}
\end{figure}

\begin{figure}[H]
    \centering
    \includegraphics[width = 0.95\textwidth]{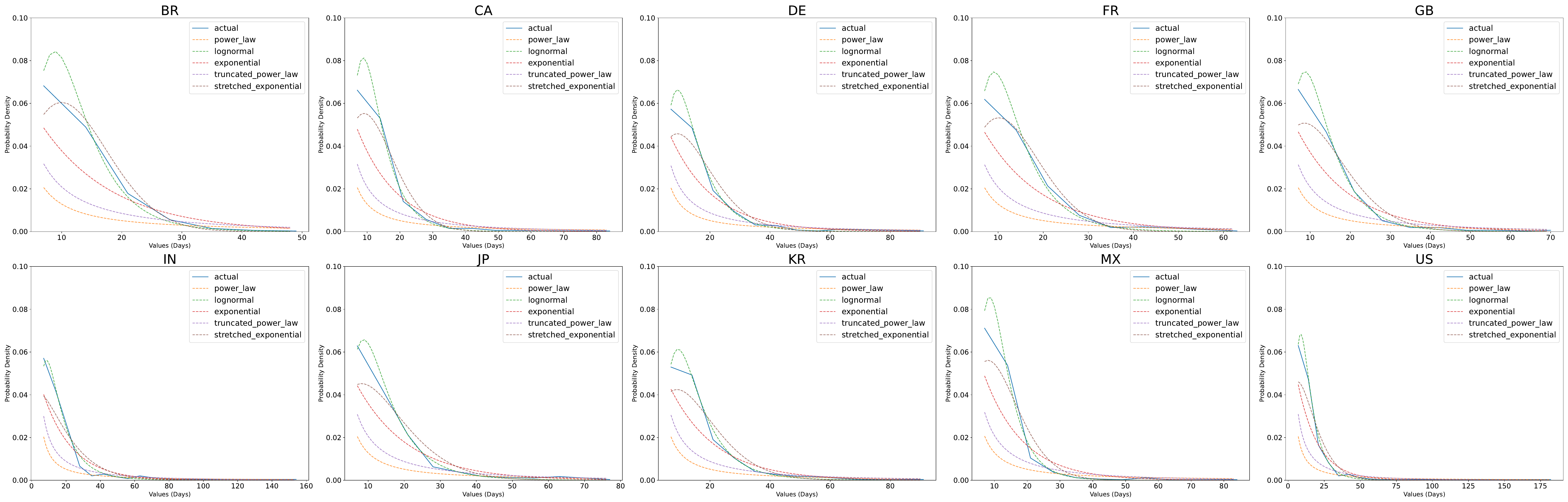}
    \caption{The fitted curve for the total survival time distribution in Netflix's film trending chart for each country. We test five model distributions (see Methods). }
    \label{fig:figs4}
\end{figure}

\begin{figure}[H]
    \centering
    \includegraphics[width = 0.95\textwidth]{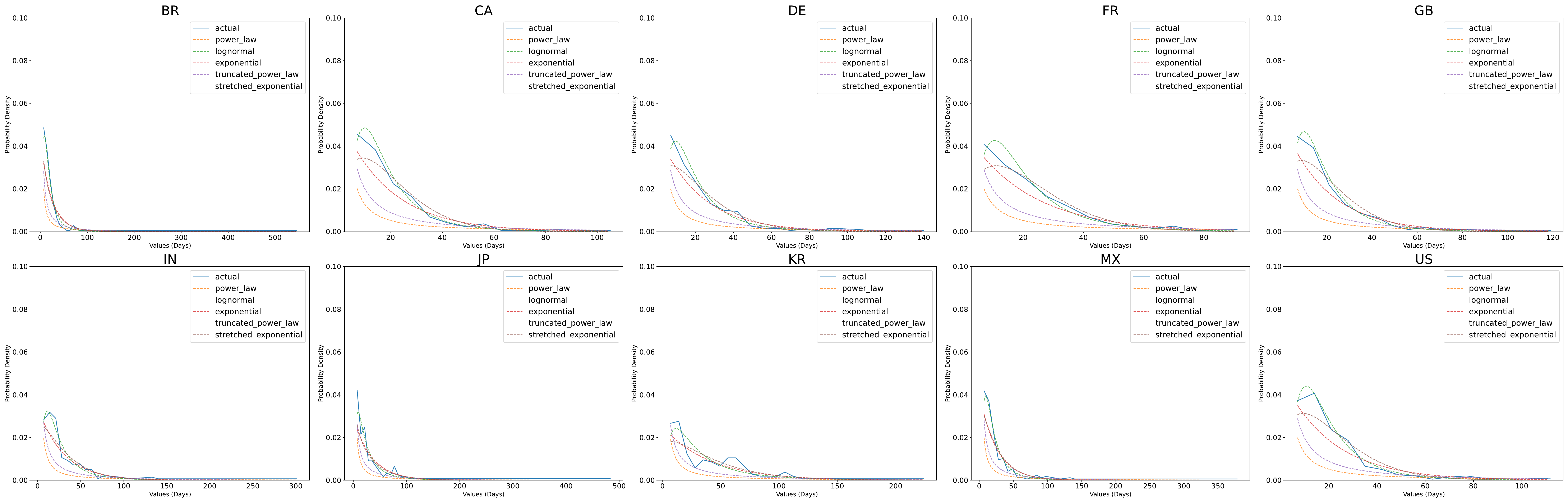}
    \caption{The fitted curve for the total survival time distribution in Netflix's TV show trending chart for each country. We test five model distributions (see Methods).}
    \label{fig:figs5}
\end{figure}

\begin{figure}[H]
    \centering
    \includegraphics[width = 0.95\textwidth]{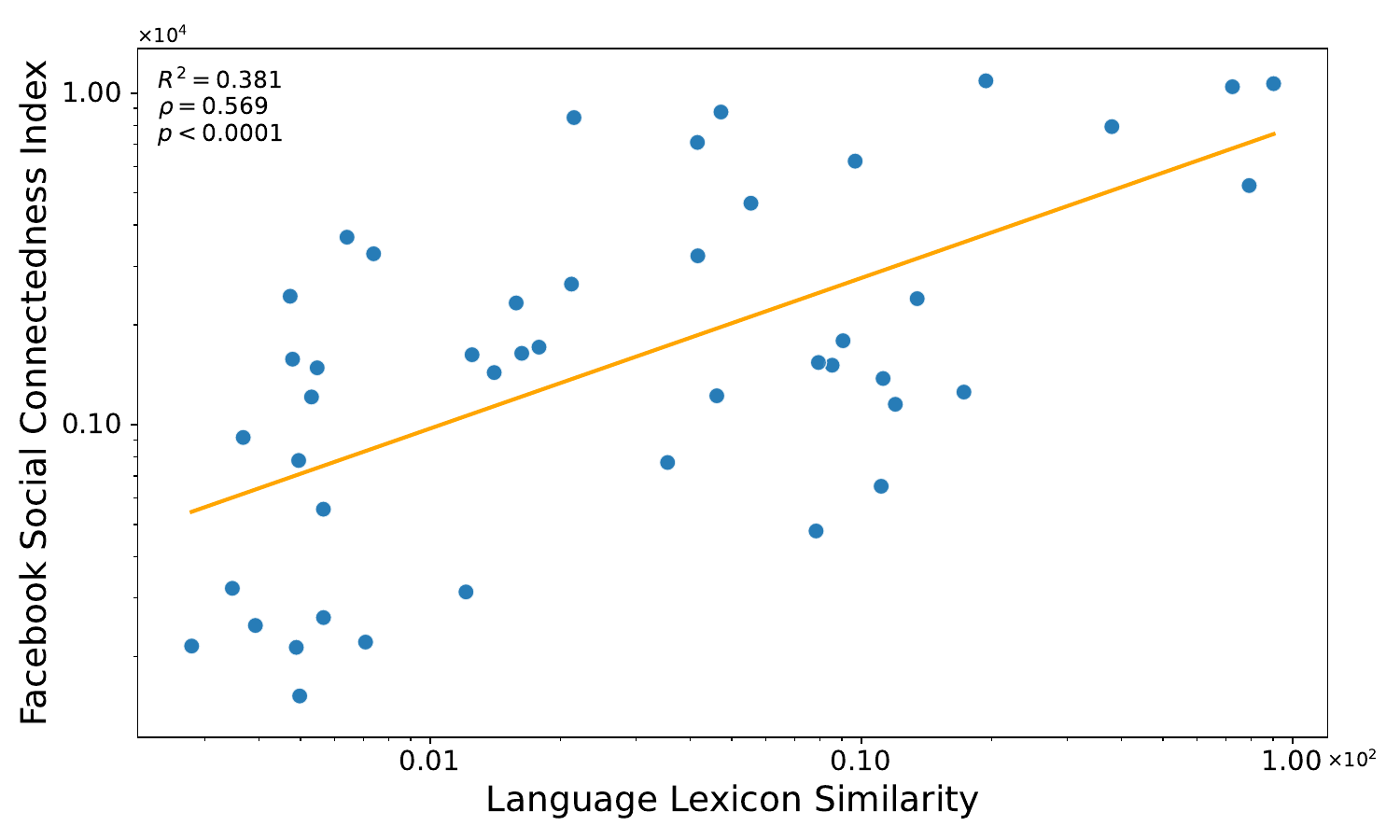}
    \caption{Correlation between two proxies of social similarity: Facebook Social Connectedness and language similarity.}
    \label{fig:figs6}
\end{figure}

\begin{figure}[H]
    \centering
    \includegraphics[width = 0.95\textwidth]{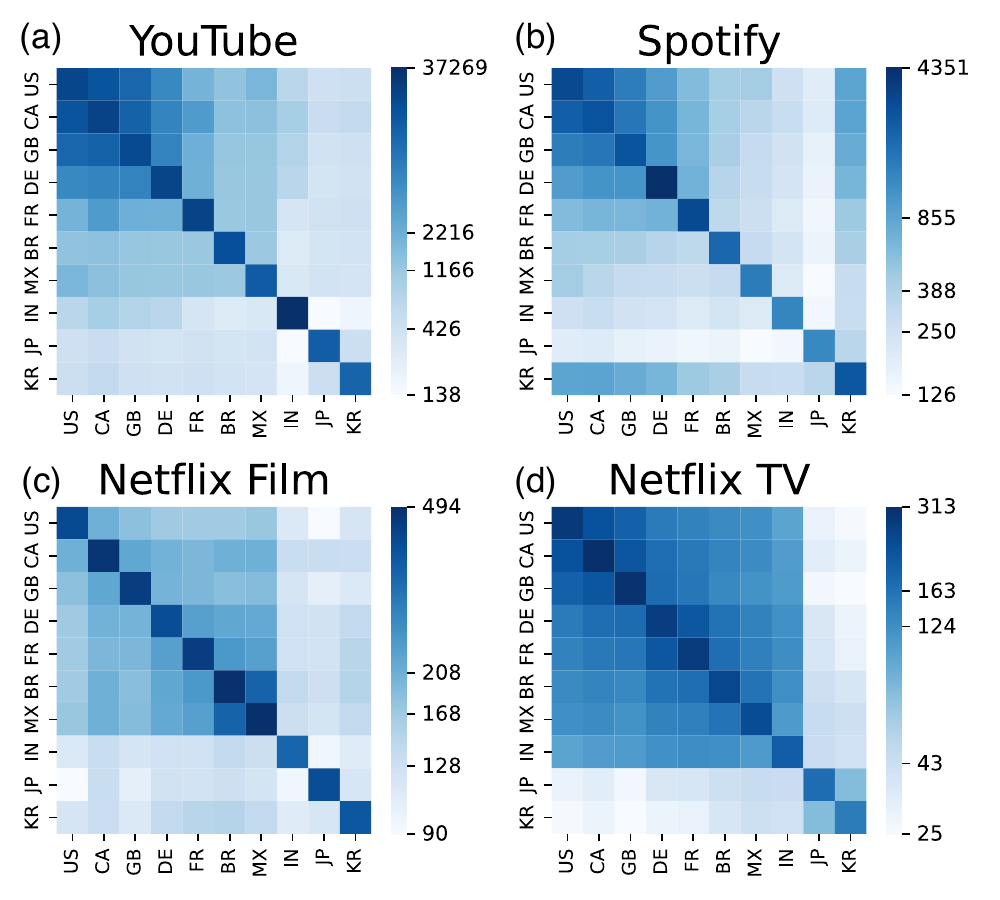}
    \caption{Numbers of shared trending content between countries with under the same period (from February 1, 2021 to February 28, 2023): (a) YouTube, (b) Spotify, (c) Netflix Film, and (d) Netflix TV show.}
    \label{fig:figs7}
\end{figure}

\begin{figure}[H]
    \centering
    \includegraphics[width = 0.95\textwidth]{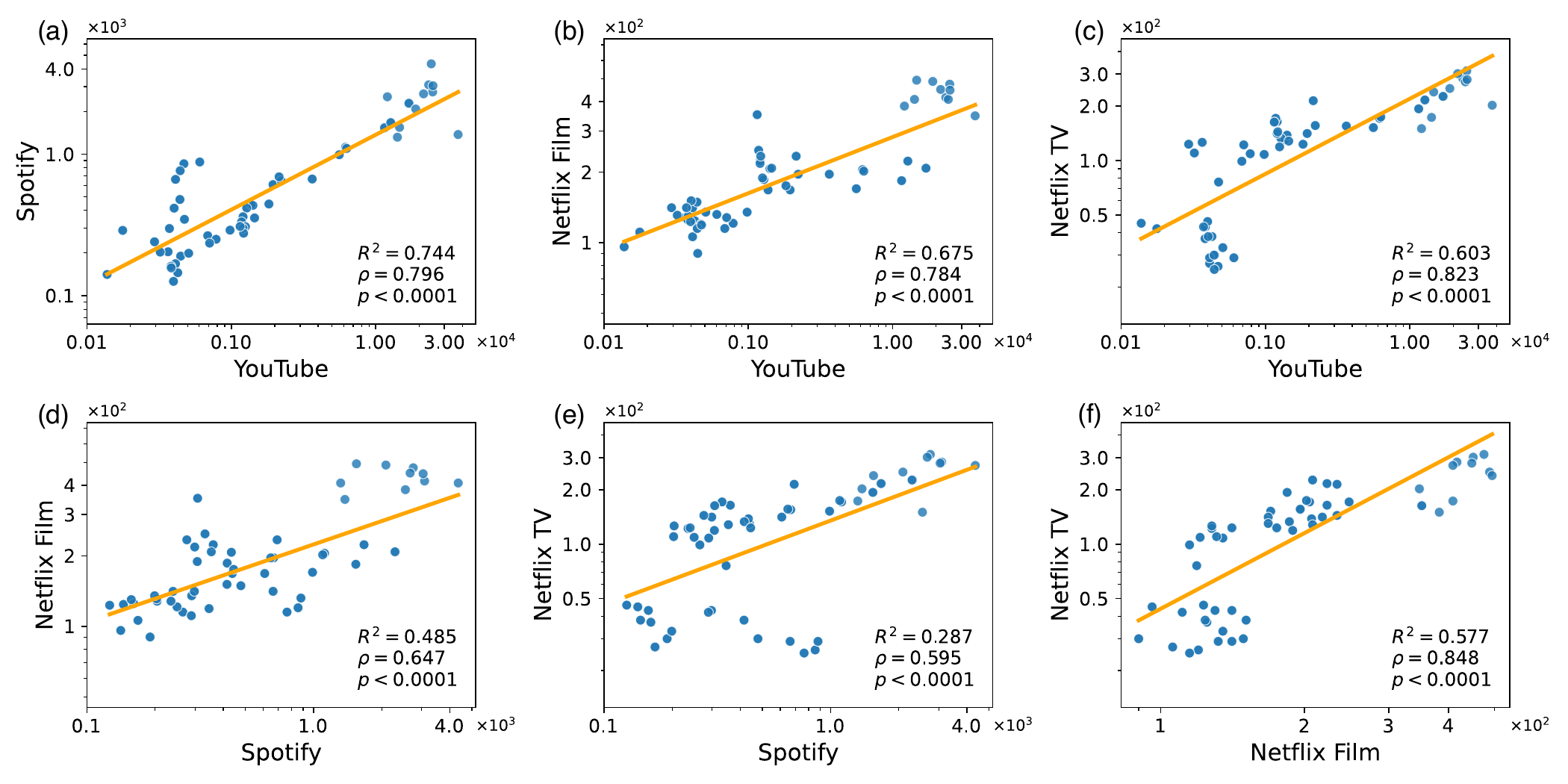}
    \caption{Cross-platform comparison for the numbers of contents between countries. Note that the time length of the three platforms are same. (from February 1, 2021 to February 28, 2023)}
    \label{fig:figs8}
\end{figure}

\begin{figure}[H]
    \centering
    \includegraphics[width = 0.95\textwidth]{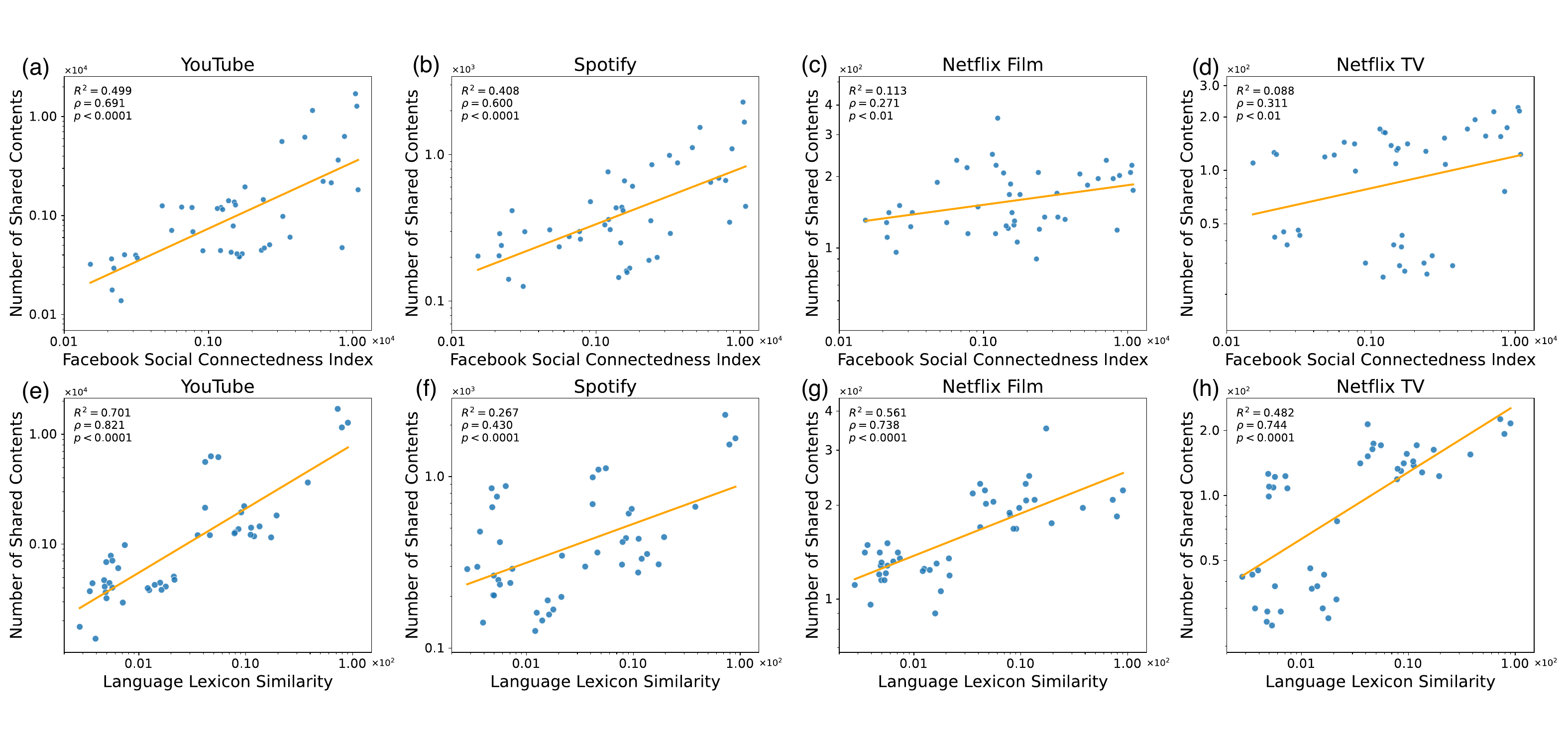}
    \caption{The correlation between the number of shared trending contents and two proxies of social similarity: (a)--(d) Facebook Social Connectedness and (e)--(h) laguange similarity. Note that the time length of the three platforms are same (from February 1, 2021 to February 28, 2023).}
    \label{fig:figs9}
\end{figure}
\pagebreak

\end{document}